\documentstyle[11pt,aaspp4]{article}
\def\gsim{\;\rlap{\lower 2.5pt
 \hbox{$\sim$}}\raise 1.5pt\hbox{$>$}\;}
\def\lsim{\;\rlap{\lower 2.5pt
   \hbox{$\sim$}}\raise 1.5pt\hbox{$<$}\;}
\newcommand\beq{\begin{equation}}
\newcommand\eeq{\end{equation}}
\def\lya{Ly$\alpha$~}
\def\rtilde{{\tilde r}}
\def\nutilde{{\tilde \nu}}
\def\nutildeobs{{{\tilde \nu}_{\rm obs}}}
\def\elltilde{{\tilde \ell}}
\def\ptilde{{\tilde p}}
\def\Itilde{{\tilde I}}
\def\Jtilde{{\tilde J}}

\def\Ktilde{{\tilde K}}
\def\Ltilde{{\tilde L}}
\def\Stilde{{\tilde S}}
\def\taumax{{\tau_{\rm max}}}
\def\rstar{{r_\star}}
\def\nustar{{\nu_\star}}
\def\Istar{{I_\star}}

\def\lyasource{{\dot N_\alpha}}  
\def\zs{{z_{\rm s}}}
\begin{document}

\title{Scattered Ly$\alpha$ Radiation Around Sources Before Cosmological
Reionization}

\author{Abraham Loeb and George B. Rybicki}
\medskip
\affil{Harvard-Smithsonian Center for Astrophysics, 60 Garden Street,
Cambridge, MA 02138}

\begin{abstract}

The spectra of the first galaxies and quasars in the Universe should be
strongly absorbed shortward of their rest-frame Ly$\alpha$ wavelength by
neutral hydrogen (HI) in the intervening intergalactic medium. However, the
Ly$\alpha$ line photons emitted by these sources are not eliminated but
rather scatter until they
redshift out of resonance and escape due to the Hubble expansion of the
surrounding intergalactic HI. We calculate the resulting brightness
distribution and the spectral shape of the diffuse Ly$\alpha$ line emission
around high redshift sources, before the intergalactic medium was
reionized.  Typically, the \lya photons emitted by a source at $z_{\rm
s}\sim 10$ scatter over a characteristic angular radius of $\sim
15^{\prime\prime}$ around the source and compose a line which is broadened
and redshifted by $\sim 10^3~{\rm km~s^{-1}}$ relative to the source.  The
scattered photons are highly polarized.  Detection of the diffuse \lya
halos around high redshift sources would provide a unique tool for probing
the neutral intergalactic medium before the epoch of reionization.  On
sufficiently large scales where the Hubble flow is smooth and the gas is
neutral, the Ly$\alpha$ brightness distribution can be used to determine
the cosmological mass densities of baryons and matter.

\end{abstract}


\section{Introduction}

Following cosmological recombination at $z\approx 10^3$, the Universe
became predominantly neutral and hence optically-thick to Ly$\alpha$
photons. The first galaxies that lit up at lower redshifts were
surrounded by a neutral intergalactic medium. Their observed spectrum
should therefore show a deep trough shortward of their rest-frame \lya
wavelength due to absorption by neutral hydrogen (HI) along the
line-of-sight, the so-called Gunn-Peterson effect (Gunn \& Peterson
1965).  For typical cosmological parameters, the optical depth at the
\lya resonance is so large that one might naively expect the damping
wing of the \lya trough to eliminate any trace of a \lya emission line
in the observed source spectrum (Miralda-Escud\'e \& Rees 1998;
Miralda-Escud\'e 1998).  However, here we point out that the \lya line
photons absorbed by intergalactic HI are subsequently re-emitted and
hence do not get destroyed\footnote{At the redshifts of interest,
$\zs\sim 10$, the low densities and lack of chemical enrichment
of the IGM make the destruction of \lya photons by two-photon decay or
dust absorption unimportant.}.  Rather, these photons scatter and
diffuse in frequency to the red of the \lya resonance due to the
Hubble expansion of the surrounding HI. Eventually, when their net
frequency redshift is sufficiently large, they escape and travel
freely towards the observer. In this paper we calculate the resulting
brightness distribution and spectral line profile of the diffuse \lya
radiation around high-redshift sources\footnote{The photons absorbed
in the Gunn-Peterson trough are also re-emitted by the IGM around the
source. However, since these photons originate on the blue side of the
\lya resonance, they travel a longer distance from the source than the
\lya line photons do, before they escape to the observer. The
Gunn-Peterson photons are therefore scattered from a larger and hence
dimmer halo around the source.  The Gunn-Peterson halo is made even
dimmer relative to the \lya line halo, by the fact that the luminosity
of the source per unit frequency is often much lower in the continuum
than in the \lya line. We therefore focus on the \lya line halo in
this paper.}.

The lack of a Gunn-Peterson trough (i.e., the detection of transmitted flux
shortward of the \lya wavelength at the source redshift) in the observed
spectra of galaxies at $z\la 5.6$ (Dey et al. 1998; Hu, Cowie, \& McMahon
1998; Spinrad et al. 1998; Weymann et al. 1998) implies that most of the
intergalactic hydrogen in the Universe was reionized before then.  Indeed,
popular cosmological models predict that reionization took place around a
redshift $z\sim 10$ (e.g., Haiman \& Loeb 1998a,b; Gnedin \& Ostriker
1997). At earlier times, the abundance of ionizing sources was small and
each of these sources produced an expanding HII region in the surrounding
intergalactic medium (IGM). The volume filling factor of HII increased
dramatically as the number of ionizing sources grew and their associated
HII regions expanded. Eventually, these HII regions overlapped and hydrogen
throughout most of the cosmic volume became ionized\footnote{Note, however,
that even after much of the IGM was ionized, the optical depth at the \lya
resonance was still substantial due to the small residual abundance of
HI. Estimates indicate that the reionized Universe provided a measurable
transmission of \lya photons only around $z\la 7$ (Haiman \& Loeb 1998;
Miralda-Escud\'e et al. 1998).} (Loeb 1997). Subsequently, the ionizing
background radiation penetrated the denser HI condensations around
collapsed objects (Miralda-Escud\'e et al. 1998; Barkana \& Loeb 1999).

The physics of reionization involves complicated gas dynamics and radiative
transfer. Numerical simulations had only recently started to incorporate
the relevant equations of radiative transfer rigorously (Abel, Norman, \&
Madau 1998).  The detection of diffuse \lya photons around high-redshift
sources offers an important empirical tool for probing the neutral IGM
before and during the reionization epoch, and can be used to test related
theoretical calculations.  In this context, one is using the \lya source as
a light bulb which illuminates the HI fog in the surrounding
IGM. Fortunately, the highest redshift galaxies are found to be strong \lya
emitters (see, e.g. Dey et al. 1998), probably due to their low dust
content.

In this paper we calculate the intensity distribution of the diffuse \lya
line in the simplest setting of a pure Hubble flow in a neutral IGM around
a steady point source. We leave more complicated configurations for future
work. In \S 2 we describe the formalism and derive analytical solutions in
the diffusion regime. In \S 3 we present a complete numerical solution to
this problem using a Monte-Carlo simulation.  The detectability of \lya
halos is discussed in \S 4. Finally, \S 5 summarizes the implications of
our results.

\section{Analytical Formalism}

Although the radiative transfer of line radiation was treated in the past
for different geometries of stationary (Harrington 1973) or moving (Neufeld
\& McKee 1988) atmospheres, no particular attention was given to the
cosmological context. As a first treatment of this involved problem, we
consider here the transfer of the \lya line in a uniform, fully neutral
IGM, which follows a pure Hubble expansion around a steady point source.
As shown later, the diffuse \lya emission around a high-redshift source
could extend over sufficiently large radii where these idealized
assumptions are indeed valid.

For newly created \lya photons the IGM is very opaque, and during
scattering events the photon frequencies are redistributed {\it
symmetrically} away from the line center due to the isotropic distribution
of thermal velocities of the hydrogen atoms.  However, as the photon
frequencies drift away from resonance and the medium becomes less opaque,
the {\it asymmetric} redshift bias imposed by the Hubble expansion of the
surrounding IGM becomes dominant.  Since the IGM is still highly opaque
when the redshift effect starts to dominate\footnote{The neutral IGM is
even colder than the microwave background prior to reionization. The
corresponding thermal velocities of hydrogen atoms at $z\la 30$ are smaller
by $\ga 3$ orders of magnitude than the Doppler velocity shift required for
the cosmological escape of resonant \lya photons ($\sim 10^{3}~{\rm
km~s^{-1}}$).}, we focus in the following only on the Hubble expansion when
calculating the {\it observed} intensity distribution around the source.
In terms of the comoving frequencies used here, this implies that the
scattering may be regarded as coherent (elastic).

We consider a source surrounded by a spherically-symmetric atmosphere
of neutral hydrogen with a linear radial velocity: $v(r)= H_{\rm s}
r$, where $H_{\rm s}$ is the Hubble expansion rate at the source
redshift, $\zs$. Let $I=I(\nu,\mu,r)$ be the specific intensity
(in ${\rm photons~cm^{-2}~s^{-1}~sr^{-1}~Hz^{-1}}$) at comoving frequency
$\nu$ in a direction $\mu=\cos\theta$ relative to the radius vector at
radius $r$, as seen by an observer comoving with the gas (i.e. in the
cosmic rest frame). Assuming isotropic coherent scattering, the
comoving transfer equation for a line at a resonant frequency $\nu_0$ is then
given by,
\begin{equation}
\mu {\partial I\over \partial r}+{(1-\mu^2)\over r} {\partial I\over
\partial \mu} +\alpha {\partial I\over \partial \nu}=
\chi_\nu\left(J-I\right) +S,
\label{eq:transfer}
\end{equation}
where $\nu$ is the frequency redshift, namely the resonant frequency $\nu_0$
minus the photon frequency; $\chi_\nu$ is the scattering opacity at $\nu$;
$J={1\over 2}\int_{-1}^{1}d\mu~I$ is the mean intensity; $S$ is the
emission function for newly created photons (in ${\rm
photons~cm^{-3}~s^{-1}~sr^{-1}~Hz^{-1}}$); and
\begin{equation}
\alpha\equiv {H_{\rm s}\nu_0\over c} = 2.7\times 10^{-13} h_0
\left[\Omega_M (1+\zs)^3 + (1-\Omega_M-\Omega_\Lambda)(1+\zs)^2
+ \Omega_\Lambda \right]^{1/2}~{\rm Hz~cm^{-1}} ,
\end{equation}
with $\Omega_M$ and $\Omega_\Lambda$ being the density parameters of matter
and the cosmological constant, and $h_0$ being the current Hubble constant,
$H_0$, in units of $100~{\rm km~s^{-1}~Mpc^{-1}}$.  Note that at
sufficiently high redshifts the value of $\alpha$ does not depend on
$\Omega_\Lambda$.  The source function on the right-hand-side of
equation~(\ref{eq:transfer}) can be written as
\begin{equation}
   S= {\lyasource \over (4\pi)^2 r^2} \delta(\nu)\delta(r),
\label{eq:source}
\end{equation}
where $\lyasource= {\rm const}$ is the steady emission rate of \lya photons
by the source (in ${\rm photons~s^{-1}}$).

In terms of our frequency variable, the \lya opacity is given by (Peebles
1993, p. 573),
\begin{equation}
\chi_\nu= \left(1-{\nu\over
\nu_0}\right)^4 {\beta \over \nu^2 + \Lambda^2 [1-\nu/\nu_0]^6/16\pi^2},
\label{eq:full_cross-section}
\end{equation}
where $\nu_0=2.47\times 10^{15}~{\rm Hz}$ is the \lya frequency,
$\Lambda=6.25 \times 10^8~{\rm s}^{-1}$ is the rate of spontaneous decay
from the $2p$ to the $1s$ energy levels of hydrogen, and
\begin{equation}
\beta\equiv  \left({3c^2 \Lambda^2\over 32 \pi^3 \nu_0^2}\right)n_{\rm HI}
= 1.5~\Omega_b h_0^2 (1+\zs)^3~{\rm cm^{-1}~Hz^2} .
\end{equation}
Here $n_{\rm HI}$ is the mean hydrogen density in the (neutral) IGM at a
redshift $\zs$, expressed in terms of the current baryonic density
parameter, $\Omega_b$, and the normalized Hubble constant, $h_0$. For
typical cosmological parameters, the mean-free-path at the line center is
negligible compared to the size of the system. Hence, a distant observer
would see only those photons which scatter to the wings of the line, where
to a good approximation the opacity scales as,
\begin{equation}
\chi_\nu\approx {\beta \over \nu^2},
\label{eq:approx_opacity}
\end{equation}
with $\beta$ being a constant due to the assumed uniformity of the IGM.
This approximate relation holds as long as the frequency shift is small,
$\nu/\nu_0\ll 1$, and is adequate throughout the discussion that follows.

Before proceeding, we should point out that the transfer problem described
by equation~(\ref{eq:transfer}) is completely analogous to that of
time-dependent monochromatic isotropic scattering, where the frequency
$\nu$ here plays the role of ``time.''  This is easily understood, since a
photon is subjected to a constant redshifting due to the Hubble expansion,
and the scattering is coherent, so its frequency is a perfect surrogate for
time.  This implies that the problem can be viewed as an initial value
problem in frequency, so the solution for a fixed frequency determines the
behavior of the solution at all other frequencies.  One should note,
however, when using this analogy that the opacity law here is a decreasing
function of ``time.''

We would now like to normalize the frequency shift and radius in the
problem by convenient scales. An appropriate frequency scale, $\nu_\star$,
is introduced by equating the \lya optical depth from the source to the
observer to unity,
\begin{equation}
\tau_\nu=\int_0^\infty {\beta\over \left(\nu_\star+\alpha r\right)^2} dr =
{\beta\over \alpha\nu_\star}=1 ,
\label{eq:tau}
\end{equation}
yielding
\begin{equation}
\nu_\star={\beta\over \alpha}= 5.6\times 10^{12} \Omega_b h_0
 \left[\Omega_M (1+\zs)^{-3} + (1-\Omega_M-\Omega_\Lambda)(1+z_{\rm
 s})^{-4}+\Omega_\Lambda (1+\zs)^{-6}\right]^{-1/2} ~{\rm Hz} .
\label{eq:nu_star}
\end{equation}
Note that for the popular values of $\Omega_b\approx 0.05$, $\Omega_M=0.4$,
$\Omega_\Lambda=0.6$ and $h_0=0.65$ (e.g., Turner 1999, and references
therein), the frequency shift of the photons which escape to infinity is
$(\nu_\star/\nu_0)\approx 4\times 10^{-3}$ at $\zs= 10$.  This
corresponds to a spectroscopic velocity shift of $(\nu_\star/\nu_0)c\sim
10^3~{\rm km~s^{-1}}$ at the source. The proper radius where the Doppler
shift due to the Hubble expansion produces the above frequency shift is
given by $r_{\star}=\nu_\star/\alpha$, or
\begin{equation}
r_\star = {\beta\over \alpha^2} = {2.1\times 10^{25}
\left({\Omega_b/\Omega_M}\right)~{\rm cm} \over
\left[1+(1-\Omega_M-\Omega_\Lambda)\Omega_M^{-1}(1+z_{\rm
s})^{-1}+(\Omega_\Lambda/\Omega_M)(1+\zs)^{-3}\right]}\approx 6.7
\left({\Omega_b\over \Omega_M}\right)~{\rm Mpc} ,
\label{eq:r_esc}
\end{equation}
The last equality was obtained for high redshifts, $\zs\gg 1$.  We
thus find that for an $\Omega_M=1$, $\Omega_\Lambda=0$ cosmology or at high
redshifts for any other cosmology, the physical distance $r_\star$ is
independent of the source redshift and depends only on the baryonic mass
fraction, $F_b\equiv \Omega_b/\Omega_M$.  This fraction can be calibrated
empirically from X-ray data on clusters of galaxies (Carlberg et al. 1998;
Ettori \& Fabian 1999) which yield a value of $F_b\approx 0.1$ for
$h_0=0.65$. Note that $\nu_\star$ was derived at the source rest frame; its
value is lower by a factor of $(1+\zs)$ in the observer's frame.

It is convenient for further developments to rescale the variables in the
problem and make them dimensionless.  Radius and frequency are rescaled
using the characteristic quantities $\rstar$ and $\nustar$, that is,
$\rtilde \equiv (r/\rstar)$ and $\nutilde \equiv (\nu/\nustar)$.  We
rescale all radiation quantities using the characteristic quantity $\Istar
= \lyasource/(\rstar^2\nustar)$, that is, $\Itilde = I/\Istar$, $\Jtilde =
J/\Istar$, etc.  We also define the rescaled source term,
\begin{equation}
\Stilde={S\rstar \over \Istar}
   = {1 \over (4\pi)^2 \rtilde^2} \delta(\nutilde)\delta(\rtilde)
\end{equation}
Equation~(\ref{eq:transfer}) can then be written in dimensionless coordinates,
\begin{equation}
\mu {\partial {\tilde I}\over \partial {\tilde r}}+{(1-\mu^2)\over {\tilde
r}} {\partial {\tilde I}\over \partial \mu} + {\partial {\tilde I}\over
\partial {\tilde \nu}}= {1\over {\tilde \nu}^2} \left({\tilde J}- {\tilde
I}\right) + {\tilde S},
\label{eq:renormalized}
\end{equation}

Next, we take the first two angular moments of
equation~(\ref{eq:renormalized}).  In addition to $\Jtilde$, these will
involve the two moments ${\tilde K}={1\over 2}\int_{-1}^{1} d\mu~\mu^2
{\tilde I}$, and ${\tilde H}={1\over 2}\int_{-1}^{1} d\mu~\mu {\tilde I}$.
In order to close the above system of equations, one needs a relation
between $\Jtilde$ and $\Ktilde$.  The closure relation is often
parameterized through the Eddington factor $f=f(\nu,r)\equiv {\tilde
K}/{\tilde J}$ (Hummer \& Rybicki 1971; Mihalas 1978).  
This parametrization yield for the comoving transfer equation
(Mihalas, Kunasz, \& Hummer 1976),
\begin{equation}
{1\over {\tilde r}^2}{\partial({\tilde r}^2 {\tilde H})\over \partial
{\tilde r}} + {\partial {\tilde J}\over \partial {\tilde \nu}}= {\tilde S}
\label{eq:first_moment}
\end{equation}
and
\begin{equation}
{\partial (f {\tilde J})\over \partial {\tilde r}}+ {(3f-1)\over {\tilde
r}}{\tilde J} + {\partial {\tilde H}\over \partial {\tilde \nu}}=-{{\tilde
H}\over {\tilde\nu}^2} .
\label{eq:second_moment}
\end{equation}

\subsection{The Extended Eddington Approximation}

We can close the above set of equations by specifying the radial and
frequency dependence of the Eddington factor $f$.  The simplest choice is
to use the Eddington approximation $f=1/3$, which is expected to be valid
at large optical depths in the medium.  However, for the moment we assume
$f$ to be a constant, but not necessarily equal to $1/3$.

For convenience, we define the rescaled variables $h\equiv {\tilde r}^2
{\tilde H}$, $j\equiv {\tilde r}^2 {\tilde J}$, and $s={\tilde r}^2 {\tilde
S}$. This yields,
\begin{equation}
{\partial h\over \partial {\tilde r}}= -{\partial j\over \partial {\tilde
\nu}}+s ,
\label{eq:hj_first}
\end{equation}
and
\begin{equation}
f{\partial j\over \partial {\tilde r}} - (1-f){j\over {\tilde r}} = 
- {\partial h\over \partial {\tilde \nu}}-{h\over{\tilde \nu}^2} .
\label{eq:hj_second}
\end{equation}
By taking the ${\tilde r}$--derivative of equation~(\ref{eq:hj_second}) and
substituting the ${\tilde \nu}$--derivative of equation~(\ref{eq:hj_first})
into it, we get
\begin{equation}
{\partial \over \partial {\tilde r}}\left(f{\partial j\over \partial
{\tilde r}} - [1-f]{j\over {\tilde r}}\right) = {\partial^2 j\over \partial
{\tilde \nu}^2} +{1\over {\tilde \nu}^{2}}{\partial j\over \partial {\tilde
\nu}} - {\partial s\over \partial {\tilde \nu}} -{s\over {\tilde \nu}^{2}}.
\label{eq:j_r_nu}
\end{equation}
If we now define the function 
\begin{equation}
g=f {\tilde r}^{-[1-f]/f}j\equiv f {\tilde r}^{[3f-1]/f}{\tilde J}.
\end{equation}
then equation~(\ref{eq:j_r_nu}) can be written as,
\begin{equation}
{\partial g\over \partial {\tilde \nu}} -  f {\tilde \nu}^2
\left({1\over {\tilde r}^{[1-f]/f}}{\partial \over \partial {\tilde r}}
{{\tilde r}^{[1-f]/f}} {\partial g \over \partial {\tilde r}} - {1\over
f}{\partial^2 g \over \partial {\tilde \nu}^2}\right) ={\tilde  \nu}^2
{\partial {\tilde s} \over \partial {\tilde \nu}} +{{\tilde s}} ,
\label{eq:wave}
\end{equation}
where $\tilde{s}=f {\tilde r}^{[3f-1]/f} {\tilde S}$.

Equation~(\ref{eq:wave}) resembles the {\it causal} diffusion equation
(Narayan, Loeb, \& Kumar 1994, and references therein), with the
substitution of frequency shift for time, and a ``diffusion'' coefficient
$D={f{\tilde \nu}^2}$. The \lya photons are expected to redshift away from
the line center at a rate which is dictated by this diffusion
coefficient. Inspection of the second-order terms provides the condition
for ``causality'' in this problem, namely ${\tilde \nu}\geq {\tilde
r}/\sqrt{f}$. This condition has a simple interpretation: the frequency
shift at any given radius is greater than the Doppler shift due to the
Hubble velocity there, since the photon could have scattered multiple times
before arriving at that radius.  Unfortunately, the line opacity is not
constant and the above diffusion equation is complicated by the frequency
dependence of the diffusion coefficient.

Equation~(\ref{eq:wave}) requires a particular choice for the value of
$f$. There are two physical regimes where the assumption of
$f=const$~~holds: (i) the diffusion regime where the optical depth is high
and $f\approx 1/3$, and (ii) the free-streaming regime where the optical
depth is low and $f\approx 1$. Note that $g={1\over 3} {\tilde J}$ in the
first regime and $g={\tilde r}^2 {\tilde J}$ in the second.  In the
diffusion regime, the number of scatterings is large and so $\nu\gg \alpha
r$ or ${\tilde \nu}\gg {\tilde r}$. This implies that the $\partial_{\tilde
\nu}^2 g$ term can be neglected relative to the $\nabla_{\tilde r}^2 g$
term in equation~(\ref{eq:wave}) and the problem is simplified
considerably.  Such a simplification is not possible in the free-streaming
case.  We therefore focus next on the diffusion regime where ${\tilde r}\ll
{\tilde \nu}\ll 1$, and $f\approx 1/3$.

\subsection{The Diffusion Solution}

In the diffusion regime, equation~(\ref{eq:wave}) reads
\begin{equation}
{\partial {\tilde J} \over \partial {\tilde \nu}} - {{\tilde \nu}^2\over 3}
{1\over {\tilde r}^2}{\partial\over \partial {\tilde r}} {\tilde
r}^2{\partial {\tilde J}\over \partial {\tilde r}}= \delta({\tilde \nu})
{\delta({\tilde r})\over (4\pi)^2 {\tilde r}^2} .
\label{eq:diffusion_regime}
\end{equation}
With the definition of a new variable, $\sigma=({{\tilde \nu}^3}/9)$, we
get the standard diffusion equation in spherical geometry,
\begin{equation}
{\partial {\tilde J}\over \partial \sigma} - \nabla^2 {\tilde J}=
   {1\over 4\pi}\delta(\sigma)\delta({{\tilde {\bf r}}}) ,
\end{equation}
where ${\bf r}$ is the radius vector, and where $\sigma$ now plays the role
of the usual time variable.  The solution to this equation for $\nu>0$
(redshifted photons) is
\begin{equation}
{\tilde J}={1\over (4\pi)^{5/2} \sigma^{3/2}} e^{-{\tilde r}^2/4\sigma} =
{1\over 4\pi} \left({9\over 4 \pi {\tilde \nu}^3}\right)^{3/2}
\exp\left\{-{9 {\tilde r}^2 \over 4 {\tilde \nu}^3}\right\} .
\label{eq:diffusion_solution}
\end{equation}

This solution satisfies the integral relation,
\begin{equation}
(4\pi)^2 \int_0^{\infty} {\tilde J} {\tilde r}^2\, d{\tilde r} = 1 ,
\label{integral_relation}
\end{equation}
which follows by performing the Gaussian integral, or directly from the
diffusion equation (\ref{eq:diffusion_regime}).  In a way this may be
viewed as a result of photon conservation, since frequency here is
analogous to time in the usual diffusion equation.

Eventually the diffusion solution becomes invalid when the opacity is small
enough to allow photons to escape to the observer.  Most of the \lya
photons escape to infinity from a smaller radius than $r_\star$ as they
scatter many times and hence redshift in frequency more than $\alpha r$ at
any given radius $r$. Even though $\nu_\star$ and $r_\star$ may not
represent the true parameters of the escaping photons,
equations~(\ref{eq:nu_star}) and~(\ref{eq:r_esc}) provide the scaling of
the exact numerical solution with model parameters, since we have already
shown that these constants can be used to normalize the basic transfer
equation~(\ref{eq:transfer}) and bring it to a scale-free form.  Next we
proceed to derive the full numerical solution for the escaping radiation in
the dimensionless variables ${\tilde \nu}$ and ${\tilde r}$.

\section{Results from a Monte-Carlo Simulation}

The lack of an analytical solution applicable to both the diffusion and
free-streaming regimes motivated us to find solutions by numerical means.
The Monte Carlo method suggested itself by its generality and ease of
formulation.  The usual disadvantage of this method of having to treat
large numbers of photons is largely irrelevant in the present case, because
all the physical parameters have been scaled out of the equations, and only
one Monte Carlo run is needed to solve the problem completely.

The Monte Carlo method follows the histories of a large number of photons
in accordance with the following rules:

\noindent
{\bf Step 1}: A newly created photon is chosen from the distribution given
by the form of the emission function $S$.  Ideally we would like this to be
a photon with $\rtilde=0$ and $\nutilde=0$.  However, because of the
singularity of the opacity, this cannot be done, so in practice we choose
our new photons with a small, but nonzero, frequency (e.g., $\nutilde =
10^{-2}$) and use the diffusion solution (\ref{eq:diffusion_solution}) to
determine a random starting value of $\rtilde$.

\noindent
{\bf Step 2}: Suppose a photon has been scattered (or has been newly
created) at a radius $\rtilde$ and with a frequency $\nutilde$.  A random
direction is then chosen from an isotropic distribution; this is most
easily done by taking $\mu = 2R-1$ where $R$ is a random deviate uniform on
the interval $(0,1)$.

To choose a step length for the photon flight, we use the fact that the
optical depth $\tau$ along the ray to the next scattering follows the
exponential distribution law, $\exp(-\tau)$.  This can be simulated by
chosing $\tau=-\ln R'$, where $R'$ is another random deviate uniform on the
interval $(0,1)$.  To convert this to distance, we compute $\tau$ as,
\begin{equation}
      \tau = \int_0^\ell \chi_\nu\, d\ell' = \int_0^{\elltilde} {d\elltilde'
                 \over (\nutilde + \elltilde')^2} = {1\over \nutilde}-
                 {1\over \nutilde+\elltilde},
\label{eq:tau_scaled}
\end{equation}
where $\ell$ is the pathlength along the ray, and $\elltilde=\ell/r_\star$.
Taking $\elltilde \rightarrow \infty$, we see that there is a maximum
optical depth to infinity,
\begin{equation}
        \taumax = {1 \over \nutilde} .
\end{equation}
If $\tau$ is greater than $\taumax$, the photon escapes and we proceed to
Step 3.  If $\tau$ is less than $\taumax$, a scattering event will
occur.  We may then solve equation~(\ref{eq:tau_scaled}) for the
pathlength,
\begin{equation}
         \elltilde = {\nutilde \tau \over \taumax- \tau}.
\end{equation}
At the new scattering site, the photon's
new radius $\rtilde'$ and  new comoving frequency $\nutilde'$ are
\begin{eqnarray}
      \rtilde' &=& ( \rtilde^2 + \elltilde^2 + 2\rtilde\elltilde \mu)^{1/2},
                  \label{eq:rinternal}\\ \nutilde'&=& \nutilde + \elltilde .
                  \label{eq:nuinternal}
\end{eqnarray}
With the substitutions $\rtilde' \rightarrow \rtilde$ and $\nutilde'
\rightarrow \nutilde$, we now repeat  Step 2 to find the positions and
frequencies of the photon at each of its successive scatterings.  This loop
is repeated until escape occurs.

\noindent
{\bf Step 3}: Escape ends the random walk for this photon.  We characterize
an escaped photon by its scaled impact parameter and its ``observed''
scaled frequency, relative to the source center,
\begin{eqnarray}
    \ptilde&=& \rtilde \sqrt{1-\mu^2},\\
      \nutildeobs &=& \nutilde - \rtilde \mu .
\label{eq:observables}
\end{eqnarray}
Here the frequency is still defined at the source rest frame and should be
divided by $(1+\zs)$ for conversion to the observer's frame.

 Steps 1, 2 and 3 are repeated for as many photons as are
necessary to get good statistical estimates for the physical quantities of
interest.

In order to relate the results of the Monte Carlo simulations to observable
quantities, histograms were constructed.  Introducing discrete sets of
impact parameters $\ptilde_i$ and ``observed'' frequencies $\nutilde_{j}$,
the final values for the escaping photons can be binned into a
two-dimensional histogram.  The ``observed'' intensity field
$\Itilde(\ptilde_{i},\nutilde_{j})$ is then estimated as proportional to
\begin{equation}
       {N_{ij} \over (2\pi \ptilde_i \Delta \ptilde_i)(\Delta \nutilde_j)},
\end{equation}
where $N_{ij}$ is the number of photons falling into the $i$,$j$ bin of
widths $\Delta \ptilde_i$, $\Delta \nutilde_j$.  Similar estimates can be
made for the intensity $\Itilde(\ptilde)$, integrated over frequencies, and
$\Itilde(\nutilde)$, integrated over the area of the plane of the sky.
Again, we define these intensities at the source frame. For conversion to
the observer's frame, $\Itilde(\ptilde_{i},\nutilde_{j})$ should be divided
by $(1+\zs)^2$, and $\Itilde(\ptilde)$ should be divided by
$(1+\zs)^3$ (since the photon phase space density $I(p, \nu)/\nu^2$
is conserved during the Hubble expansion).

Although the quantities in equations~(\ref{eq:rinternal}) and
(\ref{eq:nuinternal}) are not directly observed, it is of theoretical
interest to bin them also into a histogram in order to estimate the value
of mean intensity at interior points, based on the fact that the number of
scatterings per unit volume is proportional to the mean intensity.

A Monte Carlo run using $10^8$ photons seemed to give satisfactory results
for most of the histogram bins.  The exceptions are those at small values
of the impact parameter or radius, where the corresponding bin areas or
volumes become quite small, making the statistical errors large.

The Monte Carlo results are given in Figures 1--5.  Figure 1 shows the
intensity in the \lya line integrated over frequency,
\begin{equation}
\Itilde(\ptilde)= \int_0^{\infty} \Itilde(\ptilde,\nutilde)\,d\nutilde ,
\end{equation}
as a function of impact parameter.  The intensity has a fairly compact
central core; the characteristic impact parameter at which the intensity
has fallen to half its central value is only about $0.1 \rstar$, an order
of magnitude less than the characteristic length $\rstar$.  Because the
intensity scales inversely as the square of the characteristic impact
parameter, this means that the central intensities are two orders of
magnitude larger than a simple estimate might have indicated.  The
half-light radius $p_{1/2}$ (i.e., the radius of a circular aperture
containing half of the \lya emission) is further out, at $0.63 \rstar$, but 
at that impact parameter the intensity is already down by over an order of
magnitude from its central value.  Since the sky brightness limits the  
observational sensitivity, it may be difficult to infer $p_{1/2}$       
observationally.  However, since the scattered light is expected to be  
highly polarized (Rybicki \& Loeb 1999), its contrast relative to an    
unpolarized background could be enhanced by using a polarization filter.

The angular radius on the sky corresponding to the physical radius of
$p=0.1 \rstar$ over which the scattered \lya halo maintains a roughly
uniform surface brightness, is given by 
\begin{equation}
\theta={p\over d_{\rm A}},
\label{eq:Angular_diameter}
\end{equation}
where $d_{\rm A}$ is the angular diameter distance. In an
$\Omega_\Lambda=0$ cosmology, $d_{\rm A}=2c H_0^{-1}[\Omega_M \zs
+(\Omega_M-2)({\sqrt{1+\Omega_M \zs}}-1)]/[\Omega_M^2 (1+z_{\rm
s})^2]$ (e.g., Padmanbhan 1993), while in an $\Omega_\Lambda\ne 0$
cosmology the expression is more involved (Edwards 1972, Eisenstein 1997;
see also Fig. 13.5 in Peebles 1993). We then find for the example of
$\Omega_M=0.4$, $\Omega_\Lambda=0$, $h_0=0.65$, and $\zs=10$, that
$\theta=15^{\prime\prime}\times (p/70~{\rm kpc})$, with a slightly larger
value for a flat $\Omega_\Lambda=0.6$ cosmology. Thus, the \lya luminosity
of a typical source at $\zs\sim 10$ is expected to spread over a
characteristic angular radius of $\sim 15^{\prime\prime}$ on the sky.

Figure 2 shows the total photon emission rate (luminosity) per unit
frequency,
\begin{equation}
\Ltilde(\nutilde)= 8\pi^2 \int_0^{\infty} \Itilde(\ptilde,\nutilde)\ptilde\,
d\ptilde ,
\end{equation}
which can be used to get the observed spectral flux of photons $F(\nu)$ (in
photons cm$^{-2}$ s$^{-1}$ Hz$^{-1}$) from the entire \lya halo,
\begin{equation}
  F(\nu) = {\Ltilde(\nutilde) \over 4 \pi d_{\rm L}^2}{ \lyasource \over
             \nustar} (1+\zs)^2 ,
\end{equation}
where $\nu={\tilde \nu}\nu_\star/(1+\zs)$, and $d_{\rm L}=d_{\rm
A}(1+\zs)^2$ is the standard luminosity distance to the source. This
flux is expressed in the frame of a local observer.  The extra factor of
$(1+\zs)^2$ is due to the fact that we evaluate the photon number
flux per unit frequency rather than the energy flux -- which is used in the
usual definition of $d_{\rm L}$.  As mentioned before, the observed
intensity is
\begin{equation}
I(p,\nu)= {\Itilde(\ptilde,\nutilde)\over (1+\zs)^2}{\lyasource \over
r_\star^2 \nustar},
\end{equation}
and similarly the observed integral of the intensity over frequency,
\begin{equation}
I(p)= {\Itilde(\ptilde)\over (1+\zs)^3}{\lyasource \over
r_\star^2}.
\label{eq:intintensity}
\end{equation}

The peak of the function $\Ltilde(\nutilde)$ is seen to be redshifted by
roughly $0.44$ (recall that we define $\nutilde$ to be the {\em negative}
deviation from line center).  The full width at half maximum is $\Delta
\nutilde = 1.75$.  A linear plot of $\Ltilde(\nutilde)$ is presented in
Figure 3.  Because the shifts are small, the abscissa can be equally
interpreted as proportional to wavelength shift from line center, so Figure
3 also shows the observed lineshape as a function of wavelength.  One notes
the very extended red wing of this line, which goes approximately as
$1/\nutilde^2$.  Because of this behavior, the centroid of the line is not
well defined.

Figure 4 shows the observed intensities as a function of frequency for five
different impact parameters ($\log \ptilde =$ $0.0$, $-0.5$, $-1.0$,
$-1.5$, and $-2.0$).  The curves for $\log \ptilde=-1.5$ and $-2.0$ have
been smoothed, but still show some fluctuations due to limited Monte Carlo
sampling, especially at the smallest frequencies.  All curves share roughly the
same lineshape, and (apart from vertical scaling) differ only in the
redshift of the peak.  The peak redshift is at a minimum, $\nutilde =
0.19$, for the central rays, and increases to $0.33$ for $\ptilde = 0.1$
and to $1.4$ for $\ptilde = 1$.  Using appropriate integration, these
curves could be used to compute the expected lineshape from observations in
specific circular apertures or slits (e.g., Dey et al. 1998).

Figure 5 is mostly of theoretical interest and shows the internal mean
intensity as a function of radius for a variety of frequencies.  At
the smallest frequencies the curves are reasonably well approximated
by the diffusion solution (\ref{eq:diffusion_solution}), shown as the
dotted curves.  As $\nutilde$ increases into the free-streaming regime
($\nutilde \gtrsim 1$), the curves deviate from the diffusion
solution, and eventually become almost shock-like, with photons piling
up near the ``causality'' surface $\rtilde = \nutilde$, consistently
with the ``causal'' diffusion equation~(\ref{eq:wave}).  

One caveat concerning Figure 5 should be noted.  Many of these curves
exhibited considerable statistical error at small values of $\rtilde$,
especially for the largest values of $\nutilde$.  In these uncertain
regions the plots were completed by assuming that all curves approach
a constant value at small $\rtilde$, determined by the solution at
larger $\rtilde$.  This extrapolation procedure is certainly justified
in the diffusion limit, but its validity in the general case needs to
be investigated further.

\section{Observational Considerations}

In this section we consider the detectability of the \lya halos.  Many of
the basic properties of the early sources are unknown, and so our estimates
are meant for illustrative purposes only.

The cosmological parameters affect our estimates, but not greatly.  In this
section, we assume $\Omega_\Lambda=0$, $\Omega_M=0.4$, $\Omega_b=0.05$, and
$h_0=0.65$.  Of greater uncertainty is the \lya photon luminosity,
$\lyasource$, of high-redshift sources.  For the sake of definiteness, we
adopt the measured luminosity for a known \lya galaxy at $\zs=5.34$, which
was discovered by Dey et al.\ (1998).  To estimate the integral under the
observed spectral line profile in their Figure 3, we assume an intensity
amplitude of $\sim 5$ $\mu$Jy and a rest-frame line width of $\sim 2$ \AA\.
In our cosmological model, this source has a luminosity distance of $d_{\rm
L} = 1.64 \times 10^{29}$ cm, yielding a value of
\begin{equation}
          \lyasource = 6.4 \times 10^{53} \hbox{ s$^{-1}$} .
\label{eq:dey}
\end{equation}

Let us now assume that a candidate high-reshifted source is identified and
its reshift determined by some other unscattered lines, such as ${\rm
H}_\alpha$.  As a specific example, suppose the source is at $\zs=10$, so
that its observed \lya line falls in the infrared at $1.34 \mu{\rm m}$.
From equations (\ref{eq:nu_star}) and (\ref{eq:r_esc}) we find,
\begin{eqnarray}
        \nustar &=& 9.8\times 10^{12} \hbox{ Hz},\nonumber\\
         \rstar &=& 2.3 \times 10^{24} \hbox{ cm}= 745 \hbox{ kpc} .
\end{eqnarray}
The angular diameter distance is found to be $d_{\rm A}=3.0 \times 10^{27}$
cm, implying a typical angular size of the halo of 0.1 $\rstar/d_A = 7.7
\times 10^{-5}=16''$.

The observed integrated brightness near the center of the halo is given by
equation~(\ref{eq:intintensity}). Figure 1 shows that $\Itilde(0)=0.2$, and
so
\begin{equation}
    I_{\rm halo}(0) = 18
               \hbox{ photons cm$^{-2}$ s$^{-1}$ sr$^{-1}$} .
\label{eq:signal_halo}
\end{equation}

We suppose that a narrow filter is used to observe only in a narrow
wavelength band containing the \lya line, in order to reduce the sky
background as much as possible.  From Figure 3 we infer the Full Width at
Half Maximum of the line $\Delta \nutilde \approx 1$, implying an observing
filter width of 50 \AA\ for a source at $\zs=10$.  However, one could use
narrower filters to take advantage of the fact that the monochromatic
intensity depends jointly on frequency and impact parameter in a known way,
as shown in Figure 4.  If one wanted to concentrate only on the innermost
core of emission (out to $\ptilde = 0.1$), then a width $\Delta \nutilde =
0.5$ would be sufficient, implying a filter width of only 25 \AA.  We adopt
this narrower filter width in our estimate of the sky background.
In frequency this corresponds to $\Delta \nu = 4.2 \times 10^{11}$.

The brightness of the halo is to be compared to the sky background at 1.34
$\mu$m, which is obviously minimized in observations from space.  Using
sufficiently high resolution to eliminate contaminating point sources, the
sky brightness from space is dominated by interplanetary dust (IPD)
emission.  The local IPD contamination might be reduced by having a
spacecraft well above the ecliptic plane, but for the moment we shall
proceed conservatively and assume that the IPD is the primary sky
background.  Hauser et al.\ (1998) quote a value for the IPD brightness at
1.25 $\mu$m of $\sim 375$ nW m$^{-2}$ sr$^{-1}$.  This implies a sky
brightness within our filter band of
\begin{equation} 
I_{\rm sky} = 4.7 \times 10^{5} \hbox{ photons cm$^{-2}$ s$^{-1}$
sr$^{-1}$} .
\label{eq:sky_bgd}
\end{equation}
Thus, the halo brightness is only $\sim 4 \times 10^{-5}$ of the sky
background.  Given this low value, \lya halos would be difficult to
observe.  However, there are a number of reasons for reasonable optimism in
the long run for their detection.

The choice of \lya luminosity in equation~(\ref{eq:dey}), based on Dey et
al.\ (1998), was made purely for definiteness.  The observed radiation
intensity scales linearly with the value of $\lyasource$, and so our
estimates may be easily modified for any other value.  If the values of
$\lyasource$ for some high redshift sources were much larger (e.g. due to
higher star formation rates or lower dust content), then the difficulty in
observing the \lya halos would be eased considerably.  In this regard we
note that even if the value given in equation~(\ref{eq:dey}) is typical for
high-redshift galaxies, it is likely that there will be a power-law
distribution of observed fluxes (see Figure 2 in Haiman \& Loeb 1998b), and
some sources could have values of $\lyasource$ larger by factors of ten or
more.

In addition, the interplanetary dust emission could be much lower if the
orbit of a spacecraft such as the Next Generation Space Telescope (NGST) is
chosen so that the telescope spends some of its time outside the orbit of
the Earth or well above the ecliptic plane.  The limiting noise level in
this case is provided by the cosmic infrared background.  This background
is poorly known, but Hauser et al.\ (1998) give an upper limit at 1.25
$\mu$m of $\nu I_0 \sim 75$ nW m$^{-2}$ sr$^{-1}$, a factor of five less
than the value adopted in equation~(\ref{eq:sky_bgd}). Given this sky
background, we can calculate the minimum exposure time necessary for the
detection of the \lya halo signal.  Ignoring instrumental noise and
adopting a detector quantum efficiency close to unity (see
http:\\augusta.stsci.edu/ngst-etc/ngstetc.html, for more realistic
estimates), we can find the best signal-to-noise ratio, ${\rm S/N}$, that
is attainable after an exposure time $t$ on an NGST telescope of 8 meter
diameter,
\begin{equation}
{\rm S\over N} = 10 \left({{\dot{N_\alpha}}\over 6 \times 10^{54}~{\rm
s^{-1}}}\right) \left({t \over 10~{\rm hours}}\right)^{1/2}
\end{equation}
Hence, the halo of a source at $z_{\rm s}\la 10$ which is an order of
magnitude brighter in \lya than the Dey et al. (1998) galaxy might be
detectable.

Another circumstance favoring such observations is that the \lya halos are
highly polarized (see Rybicki \& Loeb 1999).  This polarization is highest
at the outermost radii, but is $\sim 14$\% even at the core radius of $0.1
\rstar$.  Differencing two maps in orthogonal linear polarization is
potentially a way to improve the signal-to-noise ratio.  More importantly
perhaps is that the presence of polarization would be a clear signal that
the measured halo was a \lya halo of the type described here rather than
some other, possibly instrumental, effect.

\section{Conclusions}

We have shown that \lya sources before the reionization redshift should be
surrounded by an intergalactic halo of scattered \lya photons.  These
sources are expected to appear more spatially extended in the \lya line
than they are in the continuum to the red of the line.  Our numerical
solution implies that the \lya halo has a roughly uniform surface
brightness out to an impact parameter $p\sim 0.1 r_\star\approx 70~{\rm
kpc}$ (see Fig. 1). At this impact parameter, the line is broadened and
redshifted by of order $\sim 10^3~{\rm km~s^{-1}}$ relative to the source
(Fig. 3).  These substantial broadening and redshift signatures cannot be
easily caused by galactic kinematics and hence signal the intergalactic
origin of the scattered line.

The detection of intergalactic \lya halos with the above characteristics
around sources down to a limiting redshift at which the neutral IGM ceases
to exist, can be used as a direct method for inferring the redshift of
reionization.  Alternative methods are more ambiguous, as they rely on the
detection the Gunn-Peterson damping wing which might be confused with
damped \lya absorbers along the line-of-sight (Miralda-Escud\'e 1998; see
also Haiman \& Loeb 1999 for $\zs\la 7$), the detection of the weak
damping factor of small-scale microwave anisotropies which is an integral
quantity depending also on other cosmological parameters (e.g., Hu \& White
1997; Haiman \& Loeb 1998a), or the detection of very faint (and somewhat
uncertain) spectral features in the cosmic background which is highly
challenging technologically (Haiman, Rees, \& Loeb 1997; Gnedin \& Ostriker
1997; Baltz, Gnedin, \& Silk 1998; Shaver et al. 1999).

Our calculation assumed the simplest configuration of a uniform, neutral,
IGM with a pure Hubble flow around a steady \lya source.  In popular Cold
Dark Matter cosmologies, the characteristic nonlinear mass scale of
collapsed objects at $\zs\ga 10$ is $\la 10^8~M_\odot$ (e.g., Haiman
\& Loeb 1998a).  A galaxy of total mass $10^8M_8~M_\odot$ is assembled from
a radius of $4.4 (M_8/\Omega_Mh_0^2)^{1/3}[(1+\zs)/10]^{-1}~{\rm
kpc}$ in the IGM, which is more than an order of magnitude smaller than our
inferred \lya halo radius. Similarly, the Hubble velocity at the \lya halo
radius is larger by more than an order of magnitude than the characteristic
velocity scale of nonlinear objects at these redshifts.  Hence, our
simplifying assumptions of a smooth IGM immersed in a Hubble flow are
likely to be satisfied on the \lya halo scale. Modest corrections due to
the density enhancements and peculiar velocities in the infall regions
around sources might, however, be necessary. More importantly, the
ionization effect of a bright quasar on its surrounding IGM might extend
out to the scale of interest. In such a case, the intensity distribution of
the \lya halo will depend on the spectrum and luminosity history of the
ionizing radiation emitted by the source, which determine the neutral
fraction as a function of radius around it. This ``proximity effect'' might
be important for quasars but less so for galaxies whose ultraviolet
emission is typically strongly suppressed beyond the Lyman limit due to
absorption in stellar atmospheres and in the interstellar medium.  Other
changes to the halo intensity and polarization profiles might result from
short-term variability (on $\la 10^5$ years) or anisotropic \lya emission
by the source. These complications could be easily incorporated into our
Monte Carlo approach, for particular source parameters.

Detection of the predicted \lya halo might become feasible over the next
decade either with larger ground-based telescopes or with the Next
Generation Space Telescope\footnote{NGST is the successor to the Hubble
Space Telescope which is planned for launch over the next decade. For more
details, see http://ngst.gsfc.nasa.gov/.} (NGST). For $\zs\sim 10$, the
entire \lya luminosity of a source is typically scattered over a
characteristic angular radius of $\sim 15^{\prime\prime}$.  The \lya halo
is therefore sufficiently extended to be resolved along with its tangential
polarization, as long as its brightness exceeds the fluctuation noise of
the infrared background.

Recently, five sources have been photometrically identified to have
possible redshifts of $z\ga 10$ in the Hubble Deep Field South observed by
NICMOS (Chen et al. 1998).  We emphasize that even just a narrow-band
photometric detection of the scattered \lya halos (see Fig. 1) around
sources at different redshifts would provide invaluable information about
the neutral IGM before and during the reionization epoch.  On sufficiently
large scales where the Hubble flow is smooth and the IGM is neutral, the
\lya brightness distribution can also be used to determine the values of
the cosmological mass densities of baryons and matter through
equations~(\ref{eq:nu_star}), (\ref{eq:r_esc}) and the angular diameter
distance relation. Thus, in addition to studying the development of
reionization, such observations could constrain fundamental cosmological
parameters in a redshift interval that was never probed before.

Scattering of resonance line radiation gives rise to polarization at a
level depending on the atomic physics of the transition and on the geometry
of the scattering events.  Using an extension of the Monte Carlo technique
used here, Rybicki \& Loeb (1999) showed that most photons with the largest
impact parameters suffer a nearly right-angle last scattering, and the
degree of tangential polarization is large for such photons, of order tens
of percent (asymptotically 60\%).  Measurement of this polarization could
be used as an independent check on the model parameters, and might even
provide a powerful way of identifying early objects.

\acknowledgements

We thank Chris Kochanek and Chuck Steidel for useful discussions.  This
work was supported in part by the NASA grants NAG5-7768 and NAG5-7039 (for
AL).

\begin{figure}[t]
\centerline{\epsfysize=5.7in\epsffile{ 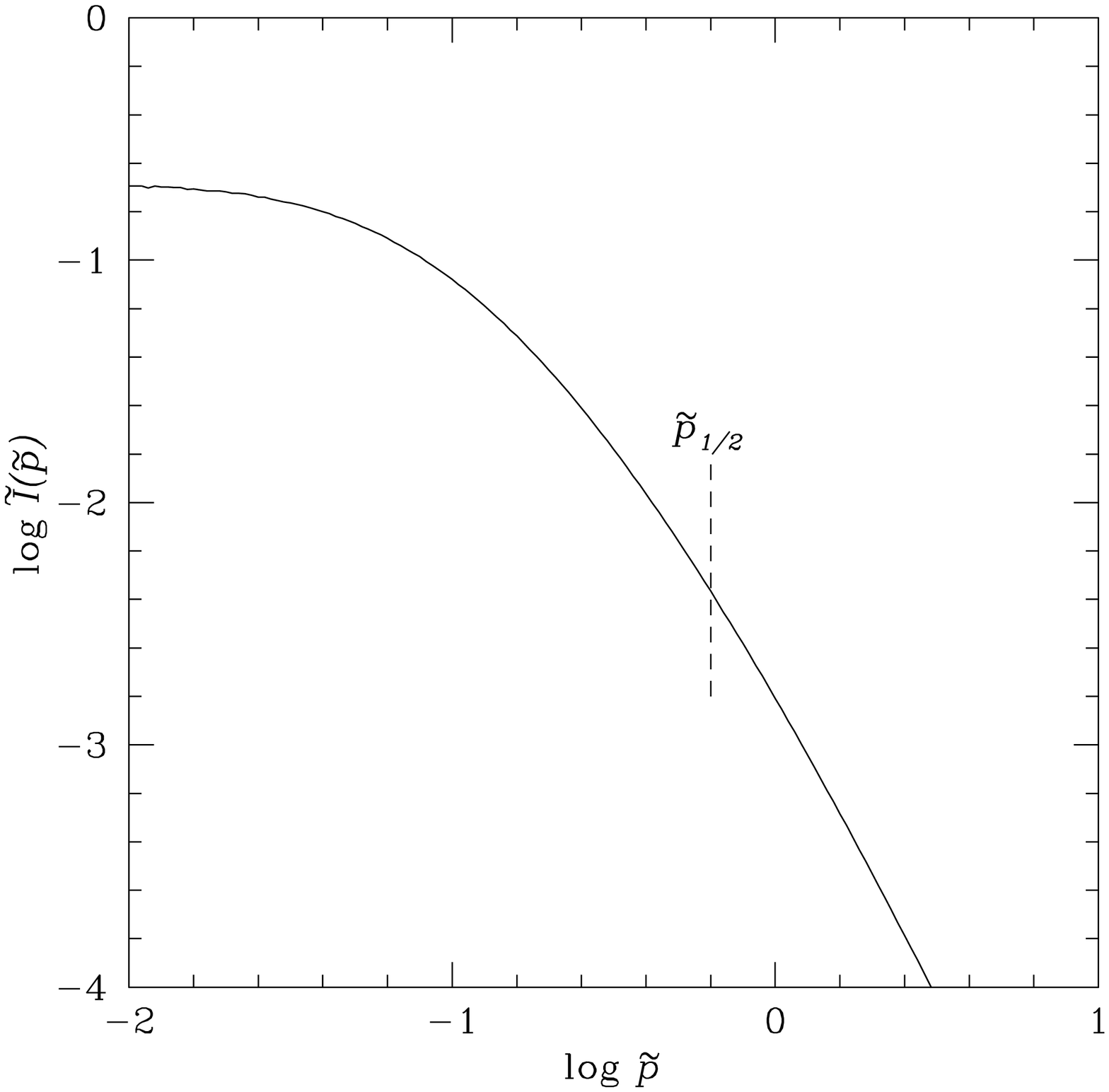 }}
\caption{Integrated intensity, ${\tilde I}({\tilde p})=\int_0^{\infty}
{\tilde I}d {\tilde \nu}$, versus impact parameter, ${\tilde p}$.  The
projected half light radius is denoted by ${\tilde p}_{1/2}$.}
\label{fig:1}
\end{figure}
\begin{figure}[t]
\centerline{\epsfysize=5.7in\epsffile{ 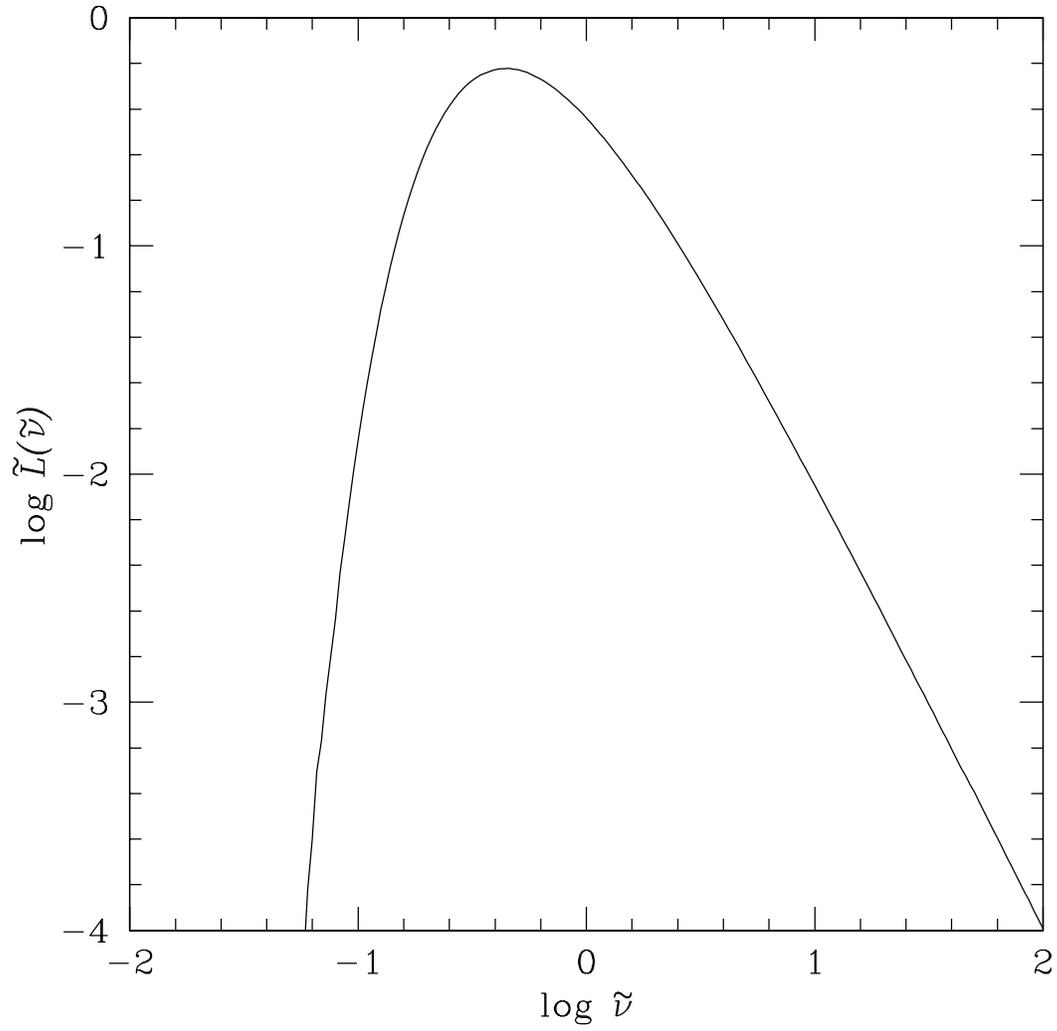 }}
\caption{Monochromatic photon luminosity, ${\Ltilde}({\tilde \nu})=8\pi^2
\int_0^{\infty} {\tilde I}~{\tilde p} d {\tilde p}$, as a function of
frequency redshift, ${\tilde \nu}$.}
\label{fig:2}
\end{figure}
\begin{figure}[t]
\centerline{\epsfysize=5.7in\epsffile{ 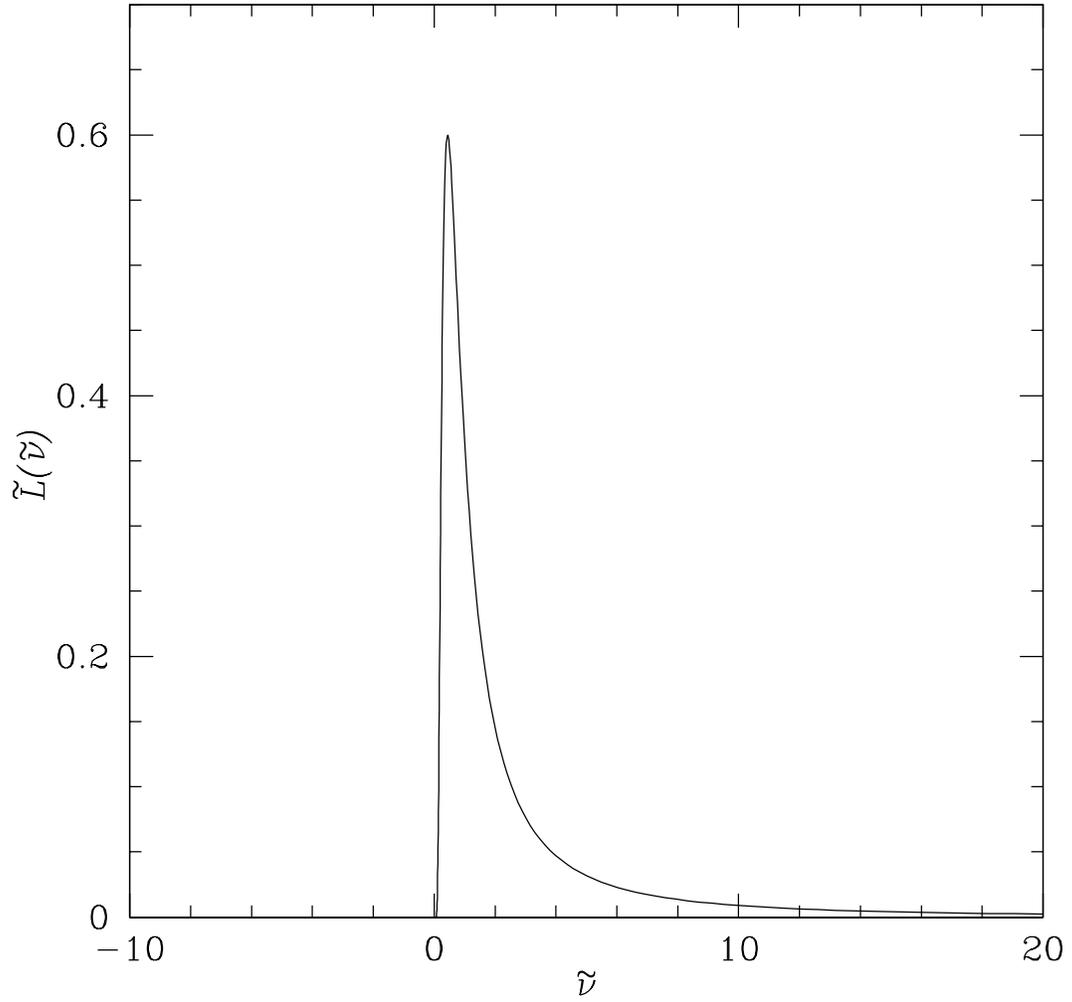 }}
\caption{Same as Figure 2, but using linear scales.}
\label{fig:3}
\end{figure}
\begin{figure}[t]
\centerline{\epsfysize=5.7in\epsffile{ 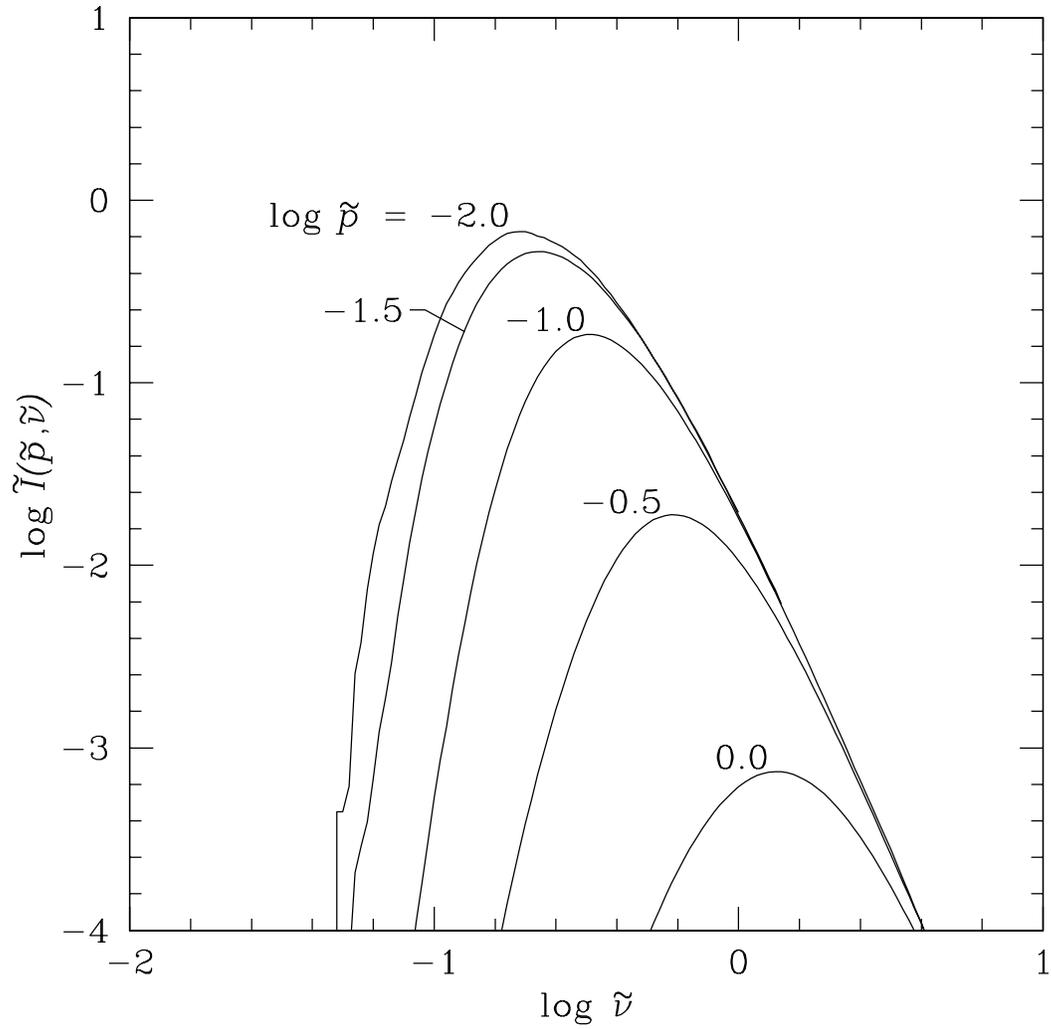 }}
\caption{\lya line profiles. Curves of intensity, ${\tilde I}({\tilde
p},{\tilde \nu})$, versus frequency shift to the red of resonance, ${\tilde
\nu}$, at different values of the impact parameter, ${\tilde p}$.}
\label{fig:4}
\end{figure}
\begin{figure}[t]
\centerline{\epsfysize=5.7in\epsffile{ 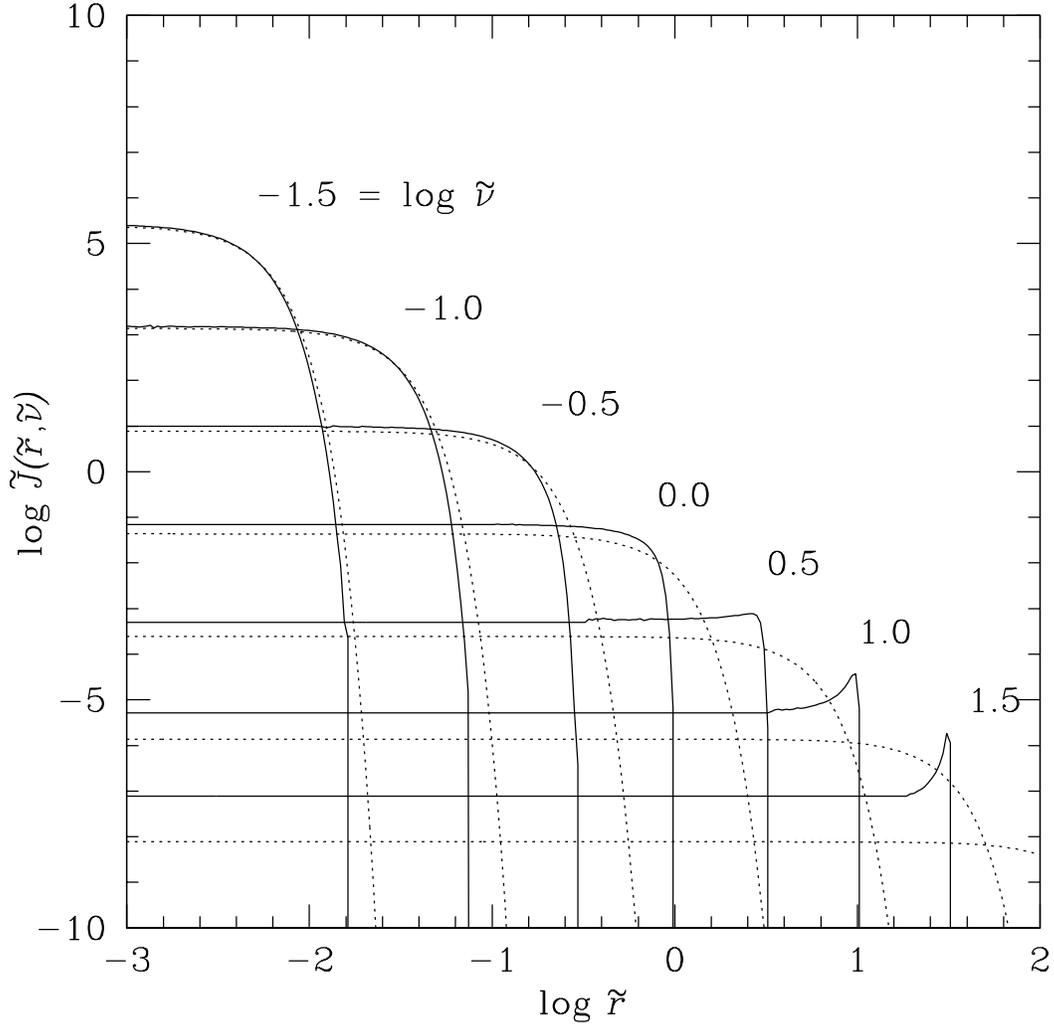 }}
\caption{Curves of the monochromatic mean intensity, ${\tilde J}({\tilde
r},{\tilde \nu})$, versus radius, ${\tilde r}$, at different values of the
frequency redshift, ${\tilde \nu}$.  Note that in agreement with the
``causal'' diffusion equation~(\ref{eq:wave}), ${\tilde r}\leq
f^{1/2}{\tilde \nu}$, where $f=1/3$ in the diffusion regime (${\tilde r}\ll
1$) and $f=1$ in the free-streaming regime (${\tilde r}\gg 1$).  The dotted
lines show the analytic solution~(\ref{eq:diffusion_solution})
for the diffusion regime.  }
\label{fig:5}
\end{figure}
\end{document}